\documentclass[nolinenumbers,amsmath,trackchanges]{aastex702}

\begin{document}

\title{Pulsar magnetospheres with stable dissipative current sheets}

\author[orcid=0000-0001-6890-4143,sname='Contopoulos']{Ioannis Contopoulos}
\affiliation{Research Center for Astronomy and Applied Mathematics, Academy of Athens,
Athens 11527, Greece}
\email[show]{icontop@academyofathens.gr}  

\author[orcid=0009-0009-2352-8709, sname='Rigas']{Spyros Rigas} 
\affiliation{Research Center for Astronomy and Applied Mathematics, Academy of Athens,
Athens 11527, Greece}
\affiliation{Department of Digital Industry Technologies, School of Science, National and Kapodistrian University of Athens, \\34400 Psachna, Greece}
\email[show]{spyrigas@dind.uoa.gr}

\author{Ioannis Dimitropoulos}
\affiliation{Research Center for Astronomy and Applied Mathematics, Academy of Athens,
Athens 11527, Greece}
\email[show]{ioannisdim.astro@gmail.com}  

\begin{abstract}

Harris-type current sheets have been studied extensively with particle-in-cell (PIC) simulations that may exhibit unrealistic characteristics when they operate at effective Lundquist numbers far below astrophysical values. Furthermore, localized studies of the pulsar equatorial current sheet generally contain no systematic magnetic-field component normal to the current sheet. Such a component develops naturally in a global pulsar magnetosphere because reconnecting field lines connect the northern and southern magnetic hemispheres. 
For realistic Lundquist numbers, the normal magnetic field may suppress the tearing instability and control the dissipation in the pulsar equatorial current sheet.
We derive the electromagnetic boundary conditions on the surface of the equatorial current sheet imposed by the surrounding ideal force-free magnetosphere and we incorporate them into a global force-free model that we solve with physics-informed neural networks (PINNs).
Preliminary solutions indicate two distinct magnetospheric configurations, a nearly ideal CKF-like state with negligible equatorial dissipation, and a strongly dissipative state resembling the solutions of \citet{Contopoulos2014} and  \citet{Cerutti2015}.
These two classes of solutions may help us understand the phenomenon of mode switching in pulsars.

\end{abstract}


\section{Introduction}

The global structure of the pulsar magnetosphere is controlled by the interaction between an almost ideal force-free plasma and a highly non-ideal equatorial current sheet extending beyond the light cylinder. Since the seminal solution of \citet{CKF1999}, hereafter CKF, the equatorial current sheet has been recognized as a key component of the pulsar electromagnetic circuit: it provides closure of the large-scale electric current and constitutes a likely site of electromagnetic dissipation, particle acceleration, and high-energy radiation. The physical conditions in this layer are therefore not a secondary detail of an otherwise fixed magnetosphere. There is a strong coupling between reconnection in the equatorial current sheet and the global magnetic-field topology, namely the distribution of electric current, the amount of open magnetic flux, the stellar spin-down luminosity and the dissipation of Poynting energy into particle energy in the current sheet
\citet{Contopoulos2007a,Contopoulos2007b,Contopoulos2007c}.
Related conclusions were reached in resistive calculations that introduced magnetospheric dissipation through prescriptions for a finite plasma conductivity \citep{Kalapotharakos2012,Li2012}. The electrostatically supported solution of \citet{Contopoulos2014} (hereafter CKK) provided a particularly clear example: the electric current closes through the equatorial sheet without a separatrix return-current layer, while a substantial fraction of the spin-down power is dissipated within a few light-cylinder radii.

Global PIC simulations have greatly advanced the study of pulsar magnetospheres by treating the plasma kinetically and following particle acceleration self-consistently. They have also reinforced the importance of the equatorial current sheet \citep[e.g.][]{Cerutti2015,Hakobyan2023,Soudais2024}. Nevertheless, present PIC simulations cannot reproduce the enormous separation between kinetic, stellar, and light-cylinder scales of real pulsars. It consequently remains difficult to distinguish robust physical features from effects controlled by numerical resolution and numerical diffusivity. This limitation is especially important when the occurrence of an instability depends directly on scales that remain unresolved.

Most local PIC studies of relativistic reconnection begin from a Harris-type sheet in which the reconnecting magnetic field reverses across the layer, possibly in the presence of a guide field. The mean magnetic-field component normal to the current sheet is then zero. A global pulsar current sheet is topologically different. Magnetic field lines approaching the sheet from the two sides originate on the central star and reconnect from the northern to the southern magnetic hemisphere. A nonzero magnetic-field component normal to the sheet therefore develops with a definite sign. As argued by \citet{Contopoulos2026}, this component is an unavoidable consequence of global magnetospheric reconnection and is absent from the initial conditions of standard local Harris-sheet calculations.

The presence of a systematic normal field changes the tearing-stability problem qualitatively. The growth of magnetic islands or plasmoids requires the bending of field lines that cross the sheet and must therefore overcome the tension associated with the normal component. A current sheet threaded by a sufficiently strong normal field can be stabilized against tearing \citep{SomovVerneta1993,Shi2021,TearingBrazil2026}. This raises a fundamental question: does the pulsar equatorial current sheet remain tearing unstable under realistic astrophysical conditions, or is the instability observed in global numerical simulations due to their low effective numerical Lundquist numbers?

In this Letter we examine the consequences of the normal magnetic field generated by global reconnection. We first combine the relevant tearing-stability criterion with the electromagnetic relations imposed by the surrounding ideal force-free magnetosphere. We then show that strong dissipation can proceed in a tearing-stable current sheet outside the light cylinder. Finally, we formulate a corresponding equatorial boundary condition and incorporate it into a global force-free model that we solve with PINNs.
We obtain two distinct classes of solutions, a nearly ideal CKF-like state with negligible equatorial dissipation, and a strongly dissipative state resembling the electrostatic configuration of CKK and the high plasma-supply PIC solution of \citet{Cerutti2015}.
We conclude by discussing the implications for the interpretation of global PIC simulations and for pulsars that switch between distinct spin-down states.

\section{Normal-field stabilization}

During reconnection, magnetic field lines from one side of a current sheet connect to field lines on the other side. In a local Harris-type configuration with complete magnetic-field reversal, with or without a guide field, no systematic large-scale normal component is imposed. Therefore, after the tearing instability produces a chain of magnetic islands, the average normal field is zero.

The situation is different in the pulsar equatorial current sheet. Once reconnection proceeds, the magnetic field lines approaching the sheet from the two sides remain anchored to the central star. Reconnection therefore joins field lines from the northern and southern magnetic hemispheres and generates a coherent normal component $\boldsymbol{B}_{\rm n}$ that crosses the current sheet perpendicularly. This global constraint is not represented in local Harris-sheet simulations. $\boldsymbol{B} _{\rm n}$ opposes the formation and growth of magnetic islands through magnetic tension. The analysis of \citet{SomovVerneta1993} showed that when
$B_{\rm n}/B_{\parallel} \ll {\cal S}^{-3/4}$, where
${\cal S}\equiv Lv_{\rm A}/\eta_{\rm m}$ is the Lundquist number of the layer, the tearing instability is not influenced by $\boldsymbol{B} _{\rm n}$. Here $B_{\parallel}$ is the magnitude of the magnetic field parallel to the current sheet immediately above or below it. In the latter definition, $L$ is the length of the current layer, $v_{\rm A}\approx c$ is the Alfven speed above and below, and $\eta_{\rm m}$ is the magnetic diffusivity in the layer. When
$B_{\rm n}/B_{\parallel} \ge {\cal S}^{-3/4}$,
the tearing instability takes place but its growth rate is diminished due to the $B_{\rm n}$. Finally, when
\begin{equation}
 \frac{B_{\rm n}}{B_{\parallel}} \gg {\cal S}^{-3/4}\ ,
 \label{eq:stability}
\end{equation}
the tearing instability is completely stabilized and only bursts of temporal spatial oscillations occur.

Let us now calculate $B_{\rm n}$ by considering the electromagnetic conditions imposed by the ideal force-free magnetosphere surrounding the equatorial current sheet.
Let the inflow reconnection speed toward the current sheet be
$v_{\rm rec} \equiv \beta_{\rm rec}c$. The corresponding reconnection electric field is parallel to the current sheet, does not reverse across it, and has magnitude
\begin{equation}
 E_{\parallel}\equiv E_{\parallel {\rm p}}=\beta_{\rm rec}B_{\parallel}^2/B_{\parallel\phi}\ ,
 \label{eq:Eparallel}
\end{equation}
where $\boldsymbol{B}_{\parallel\phi}$ is the azimuthal component of $\boldsymbol{B}_{\parallel}$.
Immediately above and below the sheet, the plasma is assumed to remain ideal and force free, and under ideal force-free steady state conditions, $\boldsymbol{E}\equiv -(r\sin\theta/R_{\rm LC})\boldsymbol{B}\times \hat{\phi}$, where $R_{\rm LC}=c/\Omega$ is the light-cylinder radius \citep{Dimitropoulos2024}, hence $\boldsymbol{E}_{\parallel}$ lies along the poloidal plane. Thus, $\beta_{\rm rec}=|\boldsymbol{E}_{\parallel}\times \boldsymbol{B}_{\parallel}|/B_{\parallel}^2=E_{\parallel}B_{\parallel\phi}/B_{\parallel}^2$.
The global corotation relation also gives
\begin{equation}
 E_{\parallel}=  \frac{r\sin\theta}{R_{\rm LC}}n_{\rm p}B_{\rm n},
 \label{eq:Bnormal_relation}
\end{equation}
where $n_{\rm p}B_{\rm n}$ is the poloidal component of $\boldsymbol{B}_{\rm n}$ \citep{Contopoulos2026}. Here, $\hat{n}\equiv n_r\hat{r}+n_\theta\hat{\theta}+n_\phi \hat{\phi}$ is the unit vector normal to the current sheet, and $n_{\rm p}\equiv (n_r^2+n_\theta^2)^{1/2}\lesssim 1$. Combining Eqs.~(\ref{eq:Eparallel}) and (\ref{eq:Bnormal_relation}) yields
\begin{equation}
 B_{\rm n}= \frac{R_{\rm LC}}{r\sin\theta\ n_{\rm p}}\,
 \beta_{\rm rec}B_{\parallel}^2/B_{\parallel\phi}\ .
 \label{eq:Bnormal}
\end{equation}
Therefore, the condition for stabilization against the tearing instability (Eq.~\ref{eq:stability}) becomes
\begin{equation}
	 \beta_{\rm rec} \gg \frac{r\sin\theta\ n_{\rm p}}{R_{\rm LC}}\ {\cal S}^{-3/4}.
 \label{eq:beta_stability}
\end{equation}
For astrophysical Lundquist numbers ${\cal S}\gg10^{5}$ \citep[e.g.][]{Zanotti2011} and for the reconnection rates  commonly measured in local relativistic PIC simulations ($\beta_{\rm rec}\sim 0.1-0.2$), Eq.~(\ref{eq:beta_stability}) is safely satisfied within at least several light cylinder radii. The pulsar equatorial current sheet should therefore be stable against the tearing instability. This conclusion creates an apparent tension with the plasmoid-dominated sheets often obtained in global PIC simulations.

\section{Two magnetospheric configurations}

In order for tearing to develop near the light cylinder, Eq.~(\ref{eq:beta_stability}) requires reconnection rates far below the canonical value $\beta_{\rm rec}\sim0.1$--$0.2$. As discussed by \citet{Contopoulos2026}, the occurrence of tearing in some global PIC simulations may instead reflect their small effective numerical Lundquist number. For example, in the global PIC simulation of \citet{Hakobyan2023}, 
\begin{equation}
 {\cal S}_{\rm num}\equiv\frac{Lc}{\eta_{\rm m}}
 \sim\frac{Lc}{d_{\rm e}^{\rm LC}\,\beta_{\rm rec}c}
 \sim10\frac{L}{d_{\rm e}^{\rm LC}}\sim 1000,
 \label{eq:Snumerical}
\end{equation}
where $L\simeq R_{\rm LC}$ denotes a characteristic length along the current sheet, and $d_{\rm e}^{\rm LC}\equiv c/\omega_{\rm pe}\simeq R_{\rm LC}/100$ is the characteristic skin depth around the light cylinder. Hence, ${\cal S}_{\rm num}^{-3/4}\simeq0.007$, and therefore a reconnection rate $\beta_{\rm rec}\sim 0.1$ is close to the numerical instability threshold for tearing and plasmoid formation to occur. Similarly, in the axisymmetric PIC simulations of \citet{HuBeloborodov2022}, $\beta_{\rm rec}=0.04$, $d_{\rm e}^{\rm LC}\simeq d_{\rm e}^{*}(R_{\rm LC}/r_*)^{3/2}\simeq R_{\rm LC}/3000\times 6^{3/2}= R_{\rm LC}/200$, 
${\cal S}_{\rm num}^{-3/4}\simeq 0.001$, which is also tearing unstable as observed. Here, $d_{\rm e}^*$ is the reported skin depth on the surface of the central star of radius $r^*\equiv R_{\rm LC}/6$.

On the other hand, the axisymmetric PIC simulations of \citet{Cerutti2015} provide an interesting contrasting case. Here,
$d_{\rm e}^{\rm LC}\simeq R_{\rm LC}/2000$, hence, for $\beta_{\rm rec}\simeq 0.1$, ${\cal S}_{\rm num}^{-3/4}\simeq0.0006$, and therefore the observed reconnection rate is safely above the threshold for stability against the tearing instability. This is probably the reason why plasmoids do not form in that simulation. Instead, the solution developed a fluctuating current sheet that remained stable against the tearing instability and dissipated approximately $30\%$ of the outgoing Poynting flux within about five light cylinder radii. This is in agreement with the prediction of \citet{SomovVerneta1993} that a current sheet that is stable against the tearing instability develops spatial oscillations. That solution was explicitly reported to resemble the electrostatically supported configuration of CKK. One possible interpretation is that the conditions realized in that simulation allowed the coherent normal field to stabilize the sheet. Whether this interpretation is correct must ultimately be established through controlled resolution and convergence studies that measure the effective diffusivity and the normal magnetic field directly. 

The comparison nevertheless motivates two physically distinct pulsar configurations. In the first, the equatorial current sheet develops at most weak tearing and negligible reconnection, and the global solution remains close to the canonical CKF state. In the second, the current sheet remains stable against tearing because of the coherent normal magnetic field. This does not imply that stabilization suppresses dissipation. In the current-sheet midplane, the raison d'etre of the normal magnetic field $\boldsymbol{B}_{\rm n}$ is the reconnection electric field $\boldsymbol{E}_{\parallel}$ (eq.~\ref{eq:Bnormal_relation}).
Particles can be accelerated wherever $E_{\parallel}>B_{\rm n}$, which is indeed possible outside the light cylinder. Thus, beyond the light cylinder, a current sheet may remain tearing stable while efficiently dissipating electromagnetic energy. Particles in the midplane undergo a drift with velocity $c\boldsymbol{E}_{\parallel}\times\boldsymbol{B}_{\rm n}/E_{\parallel}^2$ and are simultaneously accelerated along $\boldsymbol{E}_{\parallel}$. If their total speed rapidly approaches $c$, the dimensionless velocity component $\beta c$ along the electric field $\boldsymbol{E}_{\parallel}$ satisfies  $
 (B_{\rm n}/E_{\parallel})^2+\beta^2\simeq 1$,
therefore
\begin{equation}
 \beta\simeq \left[1-\left(\frac{R_{\rm LC}}{r\sin\theta\ n_{\rm p}}\right)^2\right]^{1/2}
 \label{eq:beta}
\end{equation}
\citep{Contopoulos2026}.
This expression applies only outside the light cylinder, thus identifying that region as the domain in which a stabilized sheet can accelerate particles and dissipate Poynting flux.

Consider next a current sheet  element with area $\delta A$ and thickness $h$. Poynting flux enters the layer from both sides at a rate
\begin{equation}
 2E_{\parallel}B_{\parallel\phi}\,\delta A.
\label{Poynting}
\end{equation}
The corresponding work done on the particles inside that sheet layer is
\begin{equation}
 neE_{\parallel}\beta c\,h\,\delta A,
\label{work}
\end{equation}
where $n$ is the particle number density.
We express the surface particle density $nh$ as a multiple  of the minimum number density $ \sigma_e/e$ required by the normal electric field discontinuity,
\begin{equation}
 nh = k\sigma_e/e = 2kE_{\rm n}/e\ ,
\end{equation}
where $E_{\rm n}$ is the magnitude of the normal electric field above and below the sheet, $e$ is the electron charge, and $k$ is a multiplicity factor. Under the ideal force-free conditions outside the sheet,
\begin{equation}
 E_{\rm n}=\frac{r\sin\theta}{R_{\rm LC}}n_{\rm p}B_{\parallel{\rm p}}\ ,
\end{equation}
where $B_{\parallel{\rm p}}$ is the poloidal component of the magnetic field parallel to the sheet. Finally, equating the incoming Poynting flux to the work done on the particles (eqs.~\ref{Poynting} \& \ref{work}) gives
\begin{equation}
 B_{\parallel\phi} = k\beta E_{\rm n} =  k
 \left[\left(\frac{r\sin\theta\ n_{\rm p}}{R_{\rm LC}}\right)^2-1\right]^{1/2}B_{\parallel{\rm p}}\ .
 \label{eq:field}
\end{equation}
Neither $B_{\parallel\phi}$ nor $B_{\parallel{\rm p}}$ are known a priori. They must be determined together with the global steady-state magnetospheric solution. Notice that Eq.~(\ref{eq:field})  does not involve the reconnecting electric field $E_{\parallel}$ directly. It involves it only indirectly through the global solution that yields all electromagnetic field components $\boldsymbol{E}_{\parallel}, \boldsymbol{E}_{\rm n}, \boldsymbol{B}_{\parallel}$ and $\boldsymbol{B}_{\rm n}$ in the current sheet, together with the geometry of the current sheet, namely its normal direction $\hat{n}$. It should thus be obvious to the reader that the reconnection rate in the equatorial current sheet is not determined by local current sheet microphysics as is the case in localized Harris-type PIC simulations, but self-consistently with the full magnetospheric solution. Equation~(\ref{eq:field}) generalizes in 3D Eq.~(16) of CKK by allowing a particle multiplicity $k$ above the minimum value required to support the charge of the current sheet.

The similarity between the two derivations is noteworthy because their physical starting points are different. CKK considered force balance accross the current sheet in which the electrostatic self-repulsion of the charged sheet supports it against magnetic compression. The present argument instead imposes energy balance between the Poynting flux entering the sheet and the work done on particles in its midplane. Both approaches lead to essentially the same global boundary condition.

\section{PINN solutions}

We seek representative global solutions with and without substantial dissipation in the equatorial current sheet using PINNs. The calculation extends the force-free method of \citet{Dimitropoulos2026} by imposing Eq.~(\ref{eq:field}) as a dissipative equatorial boundary condition. We solve explicitly in the hemisphere above the current sheet and use the assumed central symmetry of the electromagnetic field to determine the corresponding conditions below it. This procedure adjusts the shapes of both the equatorial sheet and the open--closed field-line boundary to achieve pressure balance accross them. 

\begin{figure*}
\centering
 \includegraphics[width=\linewidth]{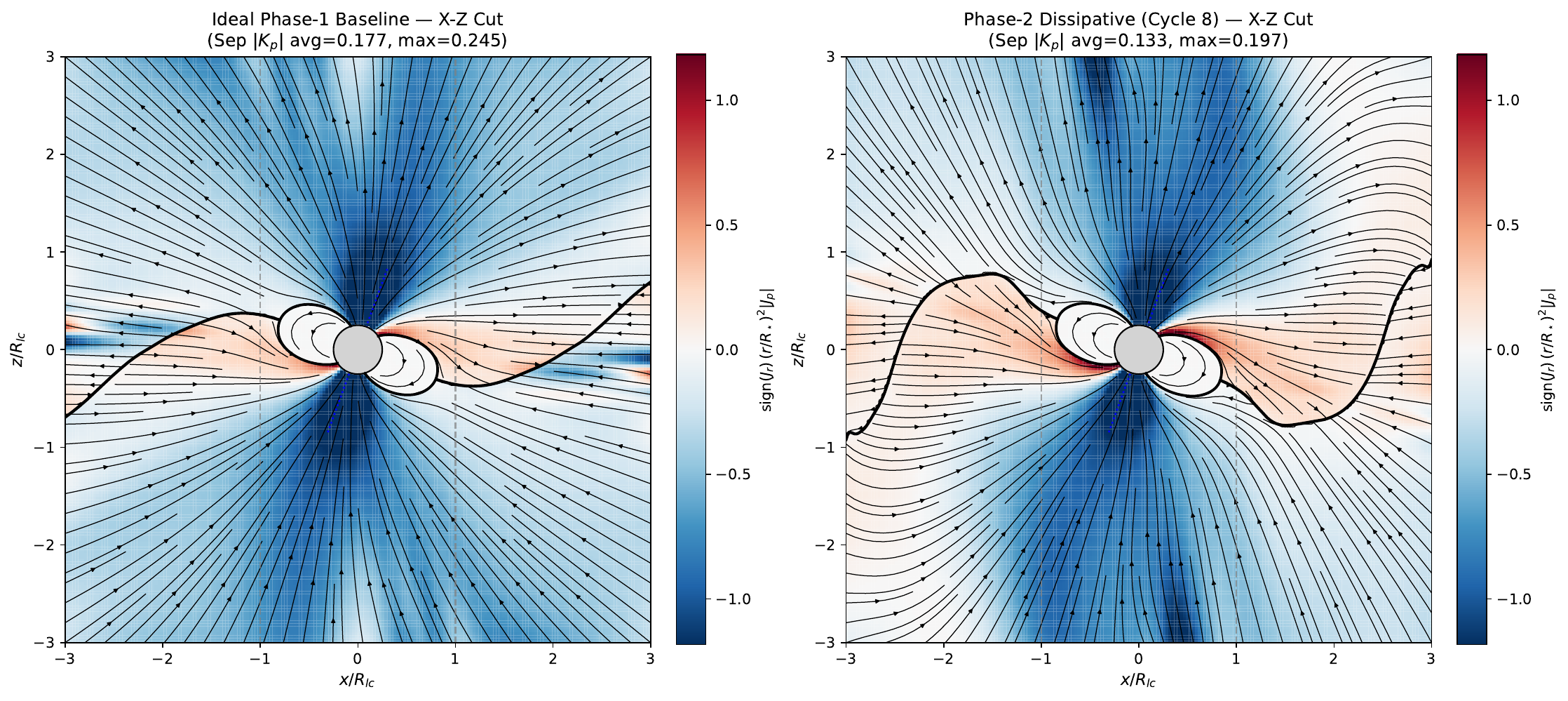}
\caption{Representative preliminary PINN solutions of the two types for $r_*=0.25 R_{\rm LC}$, inclination $20^\circ$, polar cap opening angle $36^\circ$, and multiplicity factor $k=5$. Left: CKF-like solution with negligible equatorial current sheet dissipation. Right: solution with strong dissipation in equatorial current sheet stable against the tearing instability. Color map: distribution of poloidal electric current density $\mbox{sign}(J_r)J_{\rm p}(r/R_{\rm LC})^2$ in the $x,z$ plane that contains both the rotation and magnetic axes. Thick black lines denote current sheets with infinite $J_{\rm p}$.}
\label{fig:pinn_solutions}
\end{figure*}

Preliminary calculations for pulsar inclination of $20^\circ$, polar cap opening angle of $36^\circ$, and multiplicity factor $k=5$ yield two qualitatively different types of solutions  (Figure~\ref{fig:pinn_solutions}). One remains close to the CKF configuration and exhibits negligible equatorial dissipation. The other exhibits 50\% dissipation of the spin-down power $\dot{E}$ in the equatorial current sheet between spherical radii $r=R_{\rm LC}$ and $3R_{\rm LC}$, and $\dot{E}=90\% \dot{E}_{\rm CKF}$.  It is also very interesting that, in the dissipative case, the irregular shape of the equatorial current sheet is similar to that of dissipative solutions in the literature \citep[e.g.][]{Kalapotharakos2012}.
An extended followup paper will present the numerical setup, loss functions, boundary conditions, convergence tests, magnetic-flux distributions, current functions, spin-down luminosities, and radial dissipation profiles for both types of solutions.

The coexistence of two types of solutions offers a natural framework for mode-switching pulsars. 
Transitions between a nearly ideal, higher spin-down state with almost negligible tearing instability, and a strongly dissipative, lower spin-down state with a stable equatorial current sheet could occur if changes in pair supply, current-sheet multiplicity, or effective conductivity cause the magnetosphere to switch from one global current-closure pattern to the other. A quantitative comparison with observed spin-down rates will be possible once a library of numerical solutions is completed.

\section{Conclusions}

We have argued that the tearing stability of the pulsar equatorial current sheet cannot be assessed independently of the global magnetospheric structure. In contrast to the local Harris-type configurations commonly adopted in reconnection studies, the pulsar current sheet is embedded in a magnetic topology that naturally generates a coherent field component normal to the sheet. This component follows from the connection of field lines between the northern and southern magnetic hemispheres and is therefore a global property rather than a local perturbation.

Combining the force-free relation between the reconnection electric field and the normal magnetic field with the tearing-stability criterion of \citet{SomovVerneta1993}, we find that realistic pulsar current sheets should be stable against tearing within at least a few light-cylinder radii. For astrophysical Lundquist numbers, a normal field corresponding to a nominal reconnection rate of order $0.1$ is much larger than the minimum value required for stabilization. The frequent development of plasmoids in global PIC calculations may therefore reflect, at least in part, the comparatively low effective numerical Lundquist numbers accessible to present simulations.

Stability against tearing does not imply the absence of dissipation. Outside the light cylinder, the electric field in the current-sheet midplane exceeds the normal magnetic field. Particles can then be accelerated relativistically while Poynting flux enters the sheet from both sides. Energy balance leads to the global boundary condition in Eq.~(\ref{eq:field}), which is closely related to the electrostatic support condition derived by CKK, but now includes the particle multiplicity explicitly.

The present work is not yet a complete numerical study. Our preliminary PINN calculations indicate two classes of steady global solutions. These may be relevant to mode-switching pulsars that abruptly transition between distinct spin-down regimes. The next step is to quantify the two types of solutions, test their stability against the tearing instability, and determine how their spin-down luminosity and dissipation efficiency depend on the multiplicity $k$ for various pulsar inclination angles. Furthermore, new localized relativistic PIC kinetic experiments with a global normal field component $\boldsymbol{B}_{\rm n}$ threading the reconnection layer are required to  test the central stability criterion of Eq.~(\ref{eq:stability}) and the development or absence of spatial oscillations. Such calculations are necessary before numerical reconnection with or without tearing instability can be extrapolated confidently to real pulsar magnetospheres.

\begin{acknowledgements}
The authors  acknowledge the OSCARS-funded project MEXANE, which has received funding from the European Commission's Horizon Europe Research and Innovation programme under grant agreement No. 101129751.
\end{acknowledgements}

\bibliography{Literature3}{}
\bibliographystyle{aasjournalv7.1}



\end{document}